\documentclass{emulateapj}
\usepackage{natbib}
\received{2003 March 13}
\begin{document}


\def\mbh{{$\mathcal M_{\rm BH}$}}
\def\mbhmerge{{$\mu_{\rm merge}$}}
\def\mstar{{$\mathcal M^*$}}
\def\mass{{$\mathcal M$}}
\def\rz{{$\mathcal R_{\rm {\it z}}$}}
\def\rn{{$\mathcal R_{\rm {\it 0}}$}}
\def\msol{{$\mathcal M_{\odot}$}}
\def\mr{{$M_R$}}
\def\lr{{$L_R$}}
\def\lbulge{{$L_{\rm bulge}$}}
\def\msig{{$\mathcal M_{\rm BH}$-$\sigma_*$}}
\def\mbulge{{$\mathcal M_{\rm bulge}$}}
\def\mgal{{$\mathcal M_{\rm gal}$}}
\def\mgas{{$\mathcal M_{\rm gas}$}}
\def\mstellar{{$\mathcal M_{\rm stellar}$}}
\def\mratio{{$\Gamma$}}

\shortauthors{PENG}
\shorttitle{MASS SCALING RELATIONS AND GALAXY MERGERS}

\title{How Mergers May Affect The Mass Scaling Relation Between
Gravitationally Bound Systems}

\author {Chien Y. Peng\altaffilmark{1,2}}

\altaffiltext{1}{Space Telescope Science Institute, 3700 San Martin Drive,
Baltimore, MD 21218; cyp@stsci.edu.}

\altaffiltext{2}{STScI (Giacconi) Fellow}

\begin {abstract}

Supermassive black hole (BH) masses (\mbh) are strongly correlated with galaxy
stellar bulge masses (\mbulge) and there are several ideas to explain the
origin of this relationship.  This study isolates the role of galaxy mergers
from considerations of other detailed physics to more clearly show how a
linear BH--{\it galaxy mass} relation (\mbh-\mgal) can naturally emerge
regardless of how primordial BHs were seeded inside galaxies, if the galaxy
mass function declines with increasing mass.  Under this circumstance, the
\mbh-\mgal\ relation is a passive {\it attractor} that eventually converges to
a tight linear relation because of two basic statistical effects:  a central
limit-like tendency for galaxy mergers which is much stronger for major
mergers than for minor mergers, and a convergence towards a linear relation
that is due mainly to minor mergers.  A curious consequence of this thought
experiment is that, if galaxy bulges are formed by major mergers, then merging
statistics naturally show that \mbh\ would correlate more strongly with bulge
dominated galaxies, because of stronger central-seeking tendencies, than with
disk dominated galaxies.  Even if some other physics is ultimately responsible
for causing a linear \mbh-\mbulge\ relationship, this thought experiment shows
that, counter to intuition, random merging of galaxies tends to strengthen
rather than weaken a pre-existing, linear, correlation.  This idea may be
generalized to other gravitationally bound systems (dark matter halo, compact
nuclear objects) that retain their physical identities after experiencing
mergers.

\end {abstract}


\section {INTRODUCTION}

In recent years, there have been several surprising discoveries of fundamental
scaling relations between supermassive black hole masses (\mbh) and large
scale galaxy bulge properties: stellar velocity dispersion $\sigma_*$, bulge
mass \mbulge, the profile slope of galaxies, and the inner core radius
\citep{gebhardt00a, ferrarese00, kormendy95, magorrian98, ho99, kormendy01,
graham01a, marconi03, haering04, barth05, lauer07a, greene06b,woo06}.  At
masses lower than \mbulge$\lesssim 10^{10}$\msol, the central object with
which galaxies correlate may be either intermediate mass BHs in dwarf galaxies
\citep{filippenko03, barth04, barth05, greene06b} or central massive star
clusters \citep{ferrarese06, graham07}.  The small amount of intrinsic scatter
between black holes and bulges is often interpreted to suggest a causal
connection between the two --- that the growth of one might somehow regulate
the other \citep[e.g.][]{silk98,dimatteo05}.

When did the fundamental BH scaling relationships come about?  At higher
redshifts, observations are still in the early stages, complicated in part by
the challenges of measuring the BH mass and the bulge properties in the same
galaxy.  However, recent evidence from the study of quasar host galaxies
indicates that a fundamental \mbh-\mbulge\ correlation might have been present
as early as $z\sim4$ \citep {peng06b, peng06a}.  Furthermore, there also
appears to be an evolution in the \mbh-\mbulge\ ratio by a factor of $\sim 4$
in the same studies, which points to the possibility that the BH masses may
have matured more quickly than their surrounding stellar bulge mass in the
past.  In addition, other observations that use {\sc [Oiii]} and CO emission
line widths to infer the gravitational potential of quasar bulges
\citep[]{shields03, shields06b, salviander06, salviander07, ho07} suggest
similar trends\footnote{A lack of evolution in the stellar velocity dispersion
implies that bulge mass decreases with look-back time
\citep[see, e.g, ][]{robertson06a}.}, and residual traces of evolution remain
detectable even below $z=1$ \citep{treu04, treu07, woo06}.  Despite the
general agreements observationally \citep[but, see][]{li06,borys05}, there
remain several weaknesses that still complicate the interpretation: a
significant factor of 2 systematic uncertainty in the normalization of the BH
mass calibration, the possibility that the normalization of the virial BH mass
estimate \citep{kaspi00, vester02, onken04, kaspi05, greene05a, peterson06,
vester06} may evolve with time, because of the fact that the bulge masses are
not directly measured, and the possibility that the host galaxy mass may be
biased low in high redshift quasars \citep{lauer07b} because of the steep
decline in the luminosity function of galaxies.  With regard to the latter
selection bias, it is worthwhile to note, however, that low redshift
($z\lesssim 0.3$) quasars host galaxies \citep{mclure01}, which have similarly
high \mbh\ and \mgal\ to the high redshift objects in \citet{peng06b,
peng06a}, do not exhibit a mass dependent bias even when the scatter is large.
Despite these uncertainties, the finding of an existing strong \mbh-\mbulge\
correlation at $z\gtrsim 1$ is likely to be more secure.

\begin {figure*} 
    \label{fig:distrib} 
    \plotone{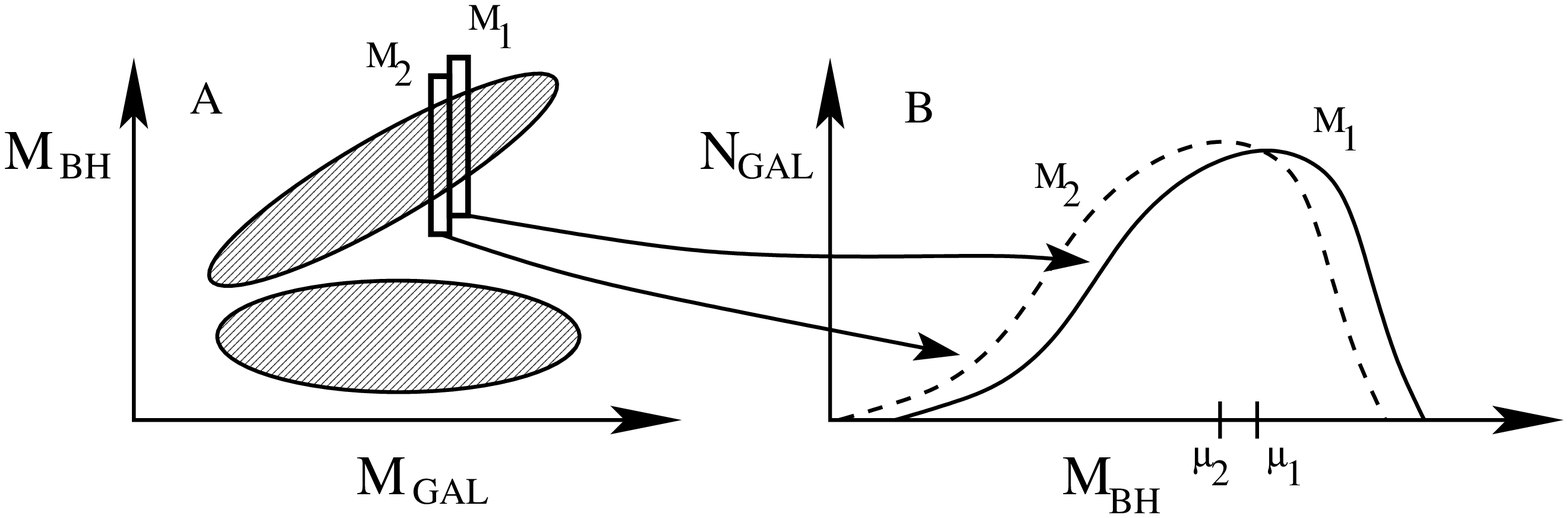}


    \figcaption{({\it a}) Two examples of initial correlations
     between \mbh\ and \mgal.  ({\it b}) An arbitrary number distribution of
     \mbh\ at each mass slice \mass$_1$ or \mass$_2$, where $\mu$ is the
     mean value of the distribution.}
\end {figure*}

Controversies about the evolution aside, it is not only puzzling that the
\mbh-bulge correlations should exist, but that they would have a small
intrinsic scatter of 0.3 dex in \mbh\, and that the correlation with the bulge
mass is practically linear \citep{marconi03, haering04, lauer07a}.  These
curious facts have received wide theoretical attention over the years and
could be explained in a number of ways.  For instance, both the \mbh-\mbulge\
and \msig\ relation can be produced by gas accretion onto a nuclear disk
\citep{burkert01, cen07} or turbulent dissipation of gas \citep{escala06}
followed by star formation, by a combination of BH accretion, galaxy mergers,
and star formation \citep{li06, kauffmann00, haehnelt00}, by gravitational
collapse of inner parts of a galaxy that forms the bulge from a rotating
isothermal sphere \citep{adams01, adams03}, by stellar capture
\citep{miralda05} in the accretion disk.  The effect of dissipation in galaxy
merging on the BH and fundamental plane scaling relations has also been
visited: \citet {ciotti01} speculated, and \citet{robertson06b} show, that
dissipational mergers can produce the \msig\ slope and maintain the
fundamental plane of elliptical galaxies.  On the other hand, whereas
dissipation{\it less} mergers of elliptical galaxies also tend to preserve the
observed fundamental plane \citep{boylan06, robertson06b}, dissipationless
mergers of disk galaxies tend to produce a relation that more closely
parallels the virial plane.

Despite the aforementioned models, the explanation that has spawned immense
activity is the theory of feedback from an active galactic nucleus (AGN).  The
AGN feedback idea rests observations that quasars typically radiate at a fixed
fraction (10\%$-$100\%) of the Eddington luminosity.  If such radiation is
produced by a massive enough BH, this energy budget is in principle sufficient
to quench star formation and terminate the BH growth itself \citep[e.g.][and
references therein]{silk98, dimatteo05, cattaneo05, springel05, hopkins07a,
hopkins06b}.  This idea has been incorporated into cosmological merger
simulations \citep[e.g.][]{granato04, fontanot06, croton06a}, and is seen to
have profound possibilities for explaining a wide array of other galaxy
evolution puzzles, including the evolution in the galaxy and quasar luminosity
functions, mass functions, star formation rates, the X-ray background, and the
bimodality of galaxy colors
\citep[e.g.  see][and references therein]{hopkins06b, hopkins06a}.

Even though AGN feedback is promising for explaining many aspects of galaxy
evolution, it is possible that some other mechanism (or mechanisms) may either
have significant influence on the scaling relation between \mbh\ and \mbulge,
or in fact be more fundamentally the cause.  It is therefore worthwhile to
look for such factors and to fully study their effects.  Dating back years
before AGN feedback physics became a popular idea, one such fundamental factor
considered by many was galaxy merging \citep[e.g.][]{kauffmann00, haehnelt00,
ciotti01, nipoti03, islam03, islam04, volonteri03}.  Even without feedback,
those simulations already find it possible to produce the BH scaling
relations, albeit to differing degrees of agreement with the observations.
However, the reason why a tight linear correlation should emerge is not
immediately apparent even when considering no other physics besides merging.
In fact, as accorded correctly by intuition, a correlation indeed {\it cannot}
arise spontaneously from chaotic combinations of galaxies and black hole
masses in general.  As will be shown below, what does make a tight correlation
emerge is the fact that the galaxy mass function decreases with increasing
mass.  This circumstance brings about a number of interesting implications
that will be explored in future studies.  In this study, the modest goal is to
show that when the focus is switched from the BH-bulge relation, to
understanding the more general BH-{\it galaxy} relation, some insights may be
gained into understanding the growth and evolution of the \mbh-\mbulge\
relation itself.

This study therefore revisits the issue of galaxy merging from the standpoint
of a thought experiment.  This toy model identifies three basic reasons for
why the \mbh-\mbulge\ relation may appear the way it does, even if the BH
masses and their host galaxy masses were completely uncorrelated initially, or
if they started out with a steep powerlaw correlation.  The over-arching
premise is that the galaxy mass function has a 
\citet{schechter76} powerlaw form, so that there is a decline in galaxy number
density with increasing mass, especially above masses \mstar.  When this
circumstance is met, it can be shown that a ``linear attractor'' and a central
limit-like tendency can work efficiently to produce a linear \mbh-\mgal\
relation, and to reduce the scatter over time.  However, while it is tempting
to generalize this result to the \msig\ relation, it is not as simple to do
because it is not clear how the stellar velocity dispersion scales during
galaxy mergers, something that depends on physics not considered in this study
\citep[e.g.  dissipational vs.  dissipationless mergers, star formation, AGN
feedback -- see][]{robertson06a, hopkins07a, dimatteo07, sijacki07}.

The main emphasis of the current study is thus only to present a pure
statistical exercise, and as such will {\it not} invoke merger trees or
external physics -- besides which many and much more sophisticated models have
already been run \citep[e.g.][]{volonteri03, islam04, granato04, fontanot06,
robertson06a, ciotti07}.  In a sense, this work examines a common thread
shared by all previous cosmological simulations.  While it is tempting to
invoke realism by introducing detailed physics from the get-go, e.g.  star
formation, BH accretion, and AGN feedback, isolating the effects of simple
statistics enables a cleaner exposition of why the convergence behavior should
be expected.  As such, the current study is not a critique of other, more
physical, models which can also explain the same correlation, or to pass
critical judgment about which is more or less relevant.  Instead, the main
message of this study is that, regardless of what other physics may ultimately
produce the \mbh-\mbulge\ relation or weaken it, an existing correlation
should strengthen if galaxies continue to merge thereafter, whether by major
or by minor mergers.  In other words, merging alone, in the absence of all
other physics, is a {\it sufficient} condition to bring about a tight, linear,
\mbh-\mbulge\ relation over time, and is always ``pulling behind the scenes''
to bring about such a correlation. The issue of interest to follow up is to
what extent this, or other as yet identified mechanisms, may matter in the
end, and to predict more realistic scatter in the \mbh-\mgal\ relation (as
opposed to only \mbh-\mbulge) under this hypothesis using more realistic
merger trees and physics.  We will address this subject in a followup study.

The following discussion will begin by presenting a heuristic view to explain
why a linear \mbh-\mgal\ relation is a natural consequence of random merging
(Section~2), followed by Monte-Carlo simulations to illustrate the effect
(Section~3), and lastly by a discussion and conclusion.  In much of the
discussion below, the relationship under consideration is more generally the
\mbh-\mgal\ relation, whereas BHs are thought to correlate more tightly with
the spheroid component of \mgal, that is \mbulge.  It will be seen that the
tighter correlation between \mbh\ and \mbulge\ is a special case of the
\mbh-\mgal\ relation and can be understood in the same framework if this
hypothesis represents the dominant mechanism by which BHs correlate with
galaxy masses.

\section {HEURISTIC PICTURES}

\subsection {The Central Limit Theorem of Galaxy Mergers}

As galaxies undergo merging, it can be shown that the scatter in the
\mbh-\mgal\ relation diminishes with increasing number of mergers as a
consequence of the central limit theorem (see Appendix).  To see this most
easily, consider first major mergers, which by definition occur between
galaxies of roughly equal mass, often defined to be within the range
\mass$_1$/\mass$_2 \le 4$ (Figure~\ref{fig:distrib}a).  Their similarity in
mass also means that the number distribution (i.e. mass function) of \mbh\ at
a specific {\it galaxy mass}, \mass$_1$, is similar to that at \mass$_2$, i.e.
they are drawn from parent distributions that have roughly the same shape
(Figure~\ref{fig:distrib}b).  However, the mean ($\mu$) of the distributions
might be offset by an amount that depends on the steepness of the
\mbh-\mgal\ correlation: the steeper the correlation
(Figure~\ref{fig:distrib}a, upper ellipse), the larger is the offset.

Furthermore, consider that {\it in the limit of no initial \mbh-\mgal\
correlation} (Figure~\ref{fig:distrib}a, lower ellipse), $\mu$ is constant
with galaxy mass.  The average of any {\it two BH} masses randomly drawn from
the parent distribution (e.g.  Fig.~\ref{fig:distrib}b) therefore has a
tendency toward the mean value $\mu$ of the parent, by the central limit
theorem.  This tendency also means that the resulting BH mass distribution
from mergers, obtained by summing BHs drawn from the same distribution, will
have a smaller log-normal scatter, $\sigma(\mbox{log}(\mu)) =
\sigma(\mu)/\mu$, than the original log-normal distribution, because the {\it
fractional mass error}, $\sigma\left<\mbox{\mbh$_{,1}$ +
\mbh$_{,2}$}\right>/\left<\mbox{\mbh$_{,1}$ +
\mbh$_{,2}$}\right>$, is smaller, which means the scatter in the \mbh\ vs.
\mgal\ relation, logarithmic on both axes, decreases.  In the instance where
the initial \mbh\ mass distribution (Figure~\ref{fig:distrib}b) is normal,
with a scatter $\sigma_{\rm{BH,init}}$, the resulting scatter of all
galaxies that have undergone one full cycle of major mergers is
$\sigma\left(\mbox{log }\mu_{\rm{BH,merge}}\right) =
\sigma \left(\mbox{log }\mu_{\rm{BH,init}}\right)/\sqrt{2}$ (see Appendix).
Therefore an ensemble of galaxies that has undergone $N_{\rm{maj}}$ mergers
should have a scatter in the \mbh-\mgal\ relation that is reduced by $\sim
2^{0.5N_{\rm{maj}}}$, compared with the initial relation.  However, a {\it
log-normal} \mbh\ distribution will have a different convergence rate.

Clearly, this central limit theorem behavior applies to a finite parent
distribution of any shape, but the size of the scatter and the rate at which
the scatter decreases both depend on the shape of the distribution and the
steepness of the initial \mbh-\mgal\ correlation.  In the situation where the
correlation between \mbh\ and \mgal\ is steep (e.g.  Fig.~\ref{fig:distrib}a,
upper ellipse) the effective cumulative mass function of BHs (Figure~1b)
residing in galaxies with \mass$_2 \le$\mass$_{\mbox{gal}}\le$ \mass$_1$ has a
wider $\sigma_{\rm{BH,init}}$ than if the \mbh-\mgal\ relation is shallow
(Fig.~\ref{fig:distrib}a, lower ellipse).  For the same reason, minor mergers
also have wider $\sigma_{\rm{BH,init}}$, as the galaxy mass differences are
larger, than do major mergers.  Therefore, galaxies that have only undergone
major mergers would produce an \mbh-\mgal\ relation that is significantly
tighter than for galaxies that have only experienced minor mergers, for the
same merger rate, and if the initial mass correlation is not flat.  This
effect is seen in the Monte-Carlo simulations below.

\begin{figure*} 
    \label{fig:correl}

    \centerline{\includegraphics[angle=0,height=8.cm,width=9.cm]{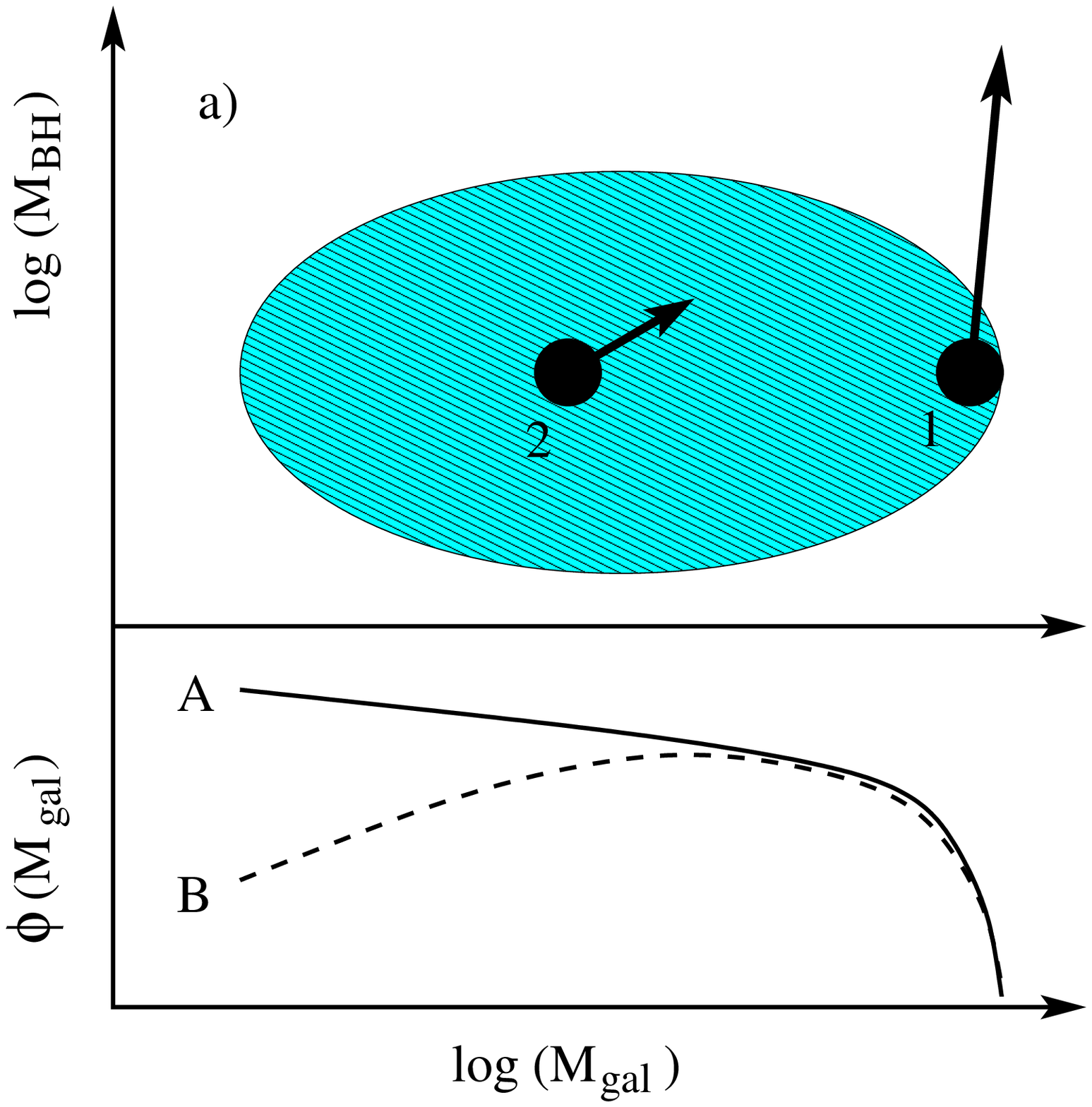}\hfill
                \includegraphics[angle=0,height=8.cm,width=7.77cm]{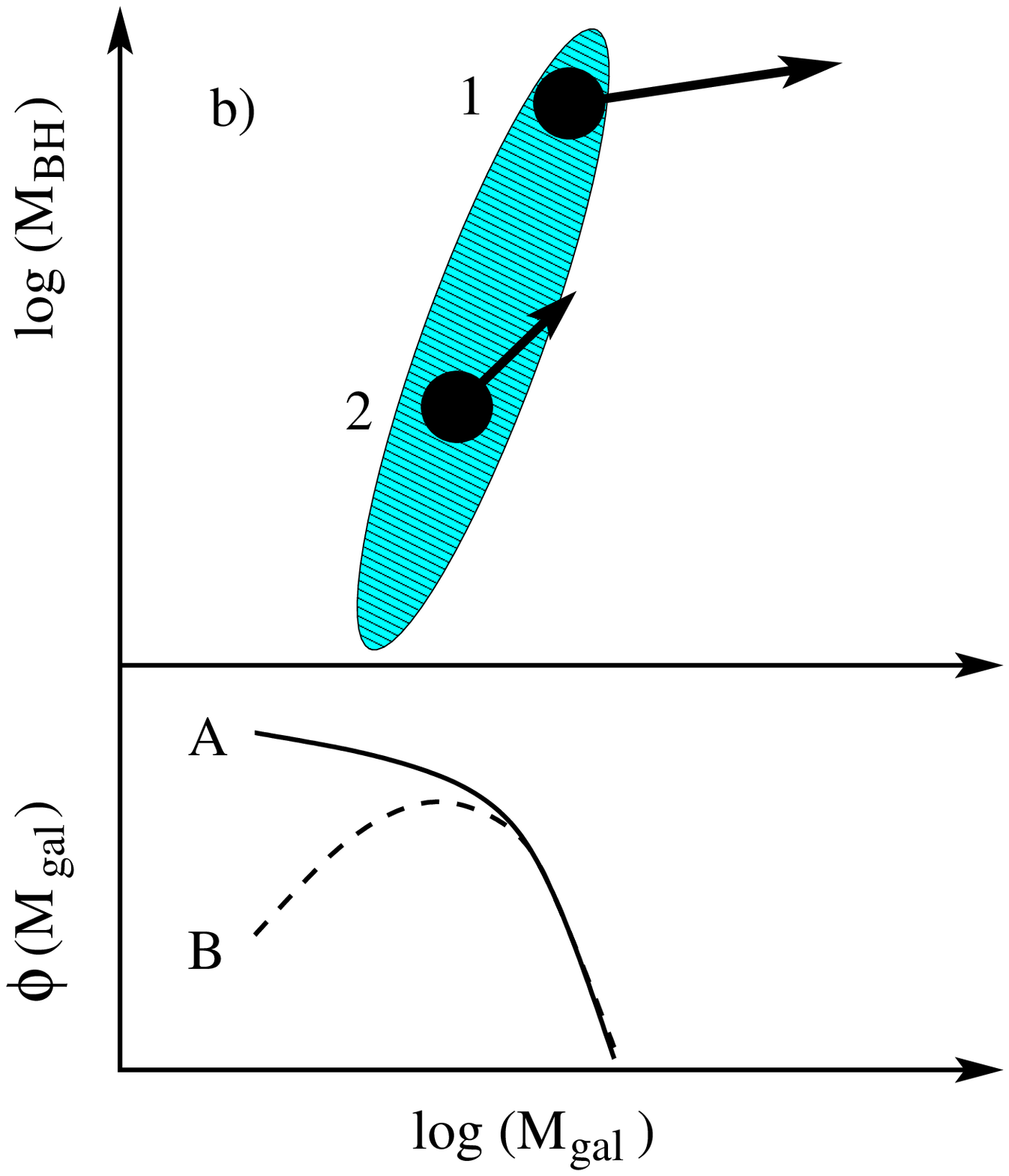}}

    \vspace{0.2in}

    \figcaption{({\it a}) No initial correlation in the \mbh-\mgal\ relation.
    A galaxy at location 1 is more likely to merge with another galaxy at a
    much lower \mgal\ than with one that is comparable to itself (due to the
    bottom heavy mass function), but roughly of comparable \mbh\ (due to
    non-correlation between \mbh-\mgal).  Thus the net evolutionary vector is
    steep.  In contrast, a galaxy at location 2 is likely to merge with
    another that is comparable in both \mbh\ and \mgal\ to itself, so the
    evolution vector is shallower.  ({\it b}) An initial, strong powerlaw,
    correlation in the \mbh-\mgal\ relation. A galaxy at location 1 is more
    likely to merge with another galaxy comparable in \mgal\ but at a much
    lower \mbh\ than itself.  Thus, the net evolutionary vector is shallow.
    In contrast, a galaxy at location 2 is likely to merge with another of
    comparable in \mbh\ and \mgal\ to itself, so the evolution vector is
    steeper.}
\end{figure*}

\subsection {Galaxy Merging From a Schechter Mass Distribution}

If the mass density of galaxies follows a Schechter powerlaw form
\citep{schechter76},

\begin{equation} 
    \Phi(\mathcal M) = \Phi_0 \left(\frac{\mathcal M}{\mathcal M^*}\right)^{\alpha+1}\rm{exp}\left(-\frac{\mathcal M}{\mathcal M^*}\right),
\end{equation}

\noindent then it is possible also to show that a linear \mbh-\mgal\
relation naturally emerges over time, so that the relation,

\begin{equation} 
    \mbox{\mbh}(z) = \mbox{\mratio}(z) \mbox{\mgal}(z)^\beta,
\end{equation}

\noindent eventually takes on $\beta = 1$.  The value of \mratio, which
locally is measured to be \mratio$(0)\sim 1/800$ for bulges
\citep[e.g.][]{haering04}, is otherwise arbitrary in the discussion below.
\mratio\ is degenerate with respect to assumptions about the initial scatter
of the \mbh-\mgal\ relation, the initial slope $\beta$, and the initial
normalization \mratio$(\infty)$, for which there are currently insufficient
observational constraints; it will not be addressed further in this study.
The other assumption used here is that the probability for two galaxies to
merge comes from Monte-Carlo sampling of a Schechter mass function (Eq. 1).
In actuality, galaxies do not merge randomly, especially at late times.
However, complete randomness is only used to facilitate the discussion, and is
not a pre-requisite, since the reasoning depends only on the fact that minor
mergers occur more frequently than major mergers.  This assumption does mean
that, depending on the relative balance of major vs.  minor mergers, the
effects described here, namely convergence toward linearity versus
central-limit behavior, may be more relevant at some epoch in time than at
others

The reason that a linear correlation emerges through galaxy mergers is
illustrated in Figure~2.  Figure~2a shows the situation in which the initial
BH and galaxy mass distributions are completely uncorrelated, so that
$\beta=0$.  The lower part of the diagram shows hypothetical mass functions
with two different ``faint end'' slopes, {\it A} and {\it B}, which will be
individually considered in the Monte-Carlo simulation below.  If there is no
correlation between the \mbh\ and \mgal, then the ratio 
\mbh/\mgal\ will be, on average, larger for low mass galaxies than for high
mass galaxies as can be seen by comparing the \mbh/\mgal\ ratio at any two
locations, for example, those labeled ``1'' and ``2.''  Therefore, as galaxies
merge, a massive object at the extreme end of the mass function, located at
position 1, {\it on average}, is more likely to merge with another having a
much larger \mbh/\mgal, thereby evolving the merger product in a steep upward
direction, as exaggerated by the vertical arrow.  In contrast, an object at
position 2 is likely to merge with objects comparable in both \mbh\ and \mgal\
to itself, so the net evolutionary vector has a shallower slope.  Therefore,
the cumulative effect of mergers along the mass spectrum is to steepen the
massive end of the \mbh-\mgal\ relation relative to the lower extreme, even as
the lower end grows in \mbh\ and \mgal\ on average.

In the other extreme, Figure~2b, if the primordial relation between the BH and
galaxy masses is steep, corresponding to $\beta \ge 1$, the opposite behavior
occurs.  Galaxies at location 1 generally have a larger \mbh/\mgal\ ratio than
galaxies that have lower mass.  Therefore, the \mbh/\mgal\ ratio for massive
galaxies would tend to decrease through mergers.  The net effect on massive
galaxies is to evolve the merger remnant more quickly to the right on average
than a lower mass galaxy at location 2.

Because of a mirror symmetry between Figures~2a and 2b, the natural
equilibrium state of the \mbh-\mgal\ relation is at $\beta = 1$, so that
further merging of galaxies would evolve remnants along the linear relation,
with a constant ratio \mratio$=$\mbh/\mgal.  Also because of the
convergence toward this ``attractor'' state the scatter in the relationship
would necessarily decrease over time through galaxy merging.

Lastly, it is worth mentioning that the presence of a break in the galaxy mass
function at \mstar\ is a sufficient, but not necessary, condition for
convergence toward linearity.  A pure powerlaw with \mstar$=\infty$ would
produce a similar behavior, however, the convergence is slower for flatter
slopes ($\alpha\rightarrow -1$).  Furthermore, while the convergence behaviors
just described is quite strong for a Schechter mass function with $\alpha=-1$,
the convergence would fail for a pure powerlaw with the same slope, because of
a lack of a break in the mass function.

\section {MONTE CARLO SIMULATIONS}

To illustrate the idea discussed above, and to quantify how quickly a linear
\mbh-\mgal\ relation might emerge, it is useful to consider several numerical
simulations for the situations shown in Figure~2.  For each of the two
scenarios, no-correlation (Fig. $2$a) and steep correlation (Fig. $2$b), it is
also instructive to consider two different initial mass functions, {\it A} and
{\it B}, shown in the lower half of Figure~2.  The two powerlaw slopes
explored below are $\alpha=-1.5$ and $\alpha=-0.5$, respectively.  These
choices are motivated by observations of the luminosity functions for high
redshift galaxies \citep[e.g.][]{gabasch06,giallongo05} under the assumption
that light traces mass.

\subsection {Simulation Set-up and Definitions}

{\it Definition of the number of major and minor mergers.} --- As implied in
Section 2, how closely a galaxy lies to the linear part of the
\mbh-\mgal\ relation, and how tight the final scatter is for an ensemble of
galaxies, will depend on the cumulative merger history.  Therefore, the most
useful way to understand the simulation results is to define the number of
major mergers, $N_{\rm{maj}}$, as the cumulative sum of all such events over
the {\it entire tree} for a given galaxy, not just in the most immediate, that
is, the main, branch.  For example, even if a galaxy has never experienced a
single major merger on the main branch in its lifetime, it could still lie
close to the final, linear part of the \mbh-\mgal\ relation because the {\it
progenitors} of the main branch, and their progenitors, and so forth, could
have experienced a number of major mergers.

{\it Evolution of the mass function.} --- One issue to consider is how the
galaxy mass function might evolve, and whether the path of evolution might
affect the final convergence.  The effect of galaxy masses growing with time
is to both increase \mstar\ and steepen the ``faint end slope'' ($\alpha$) of
the Schechter function.  As galaxy populations grow in mass and number
density, the rate of change in \mstar\ and $\alpha$ would affect the rate of
convergence to the final \mbh-\mgal\ normalization, slope, and scatter.
Predicting the rate of change in the \mbh-\mgal\ relation requires realistic
merger trees and accounting for other detailed physics such as feedback, which
will be addressed in a followup study.  For the current purposes of showing
that convergence toward a linear \mbh-\mgal\ relation does naturally occur, it
suffices to consider two scenarios: a {\it replenishment scenario}, in which
the Schechter mass function is continuously, and randomly, replenished as
galaxies merge, and a {\it depletion scenario}, in which no new galaxies are
formed to take the place of those that have merged.  Combined with
considerations about the initial mass function slopes, the simulations will
have covered the gamut of sensible possibilities and conditions that may be
present at early and late cosmic epochs.

{\it Initial scatter of the \mbh-\mgal\ relation and the initial mass function
\mstar.} --- In the simulation, the distribution of galaxies is first drawn
randomly from the Schechter mass function, after which a BH mass is assigned,
following Equation~2, by drawing from a log-normal Gaussian distribution with
a generous Gaussian dispersion of $\sigma_{\rm BH}=2$ dex, i.e.  a scatter of
a factor 100 in mass.  The exact choice of the initial scatter is directly
proportional to, but otherwise only partially determines, the final scatter in
the \mbh-\mgal\ relation.  Other factors that determine the final scatter
depend on how long the simulation runs, and on the initial value of
\mstar.  Currently, there are some observational constraint on the rate of
mergers, which will be considered in a followup study.  In this study, the
results are merely normalized arbitrarily to match the final
\mbh-\mgal\ relation observed today.

{\it The simulation ``clock.''} --- The progress of time is not well defined
in Monte-Carlo simulations, so it is useful to define {\it merging cycles} for
the sake of keeping track of the simulation progress.  Each full cycle is
defined as being complete after the number of merger events equals the number
of galaxies present at the beginning of that particular cycle.  Galaxies that
are produced or merged retain their states for the following cycle.  Because
some galaxies may merge multiple times by being drawn repeatedly, not all
galaxies will be involved in mergers after each full cycle.  The exact
definition of the simulation clock is unimportant, as it is only the relative
number of major vs. minor mergers on average that determines the degree of
convergence toward a linear \mbh-\mgal\ relation, where a major merger is
defined as having a mass ratio of at most 4:1.


\begin{figure*} 
    \label {fig:nocorr}
    \centerline{\includegraphics[angle=0,width=9.cm]{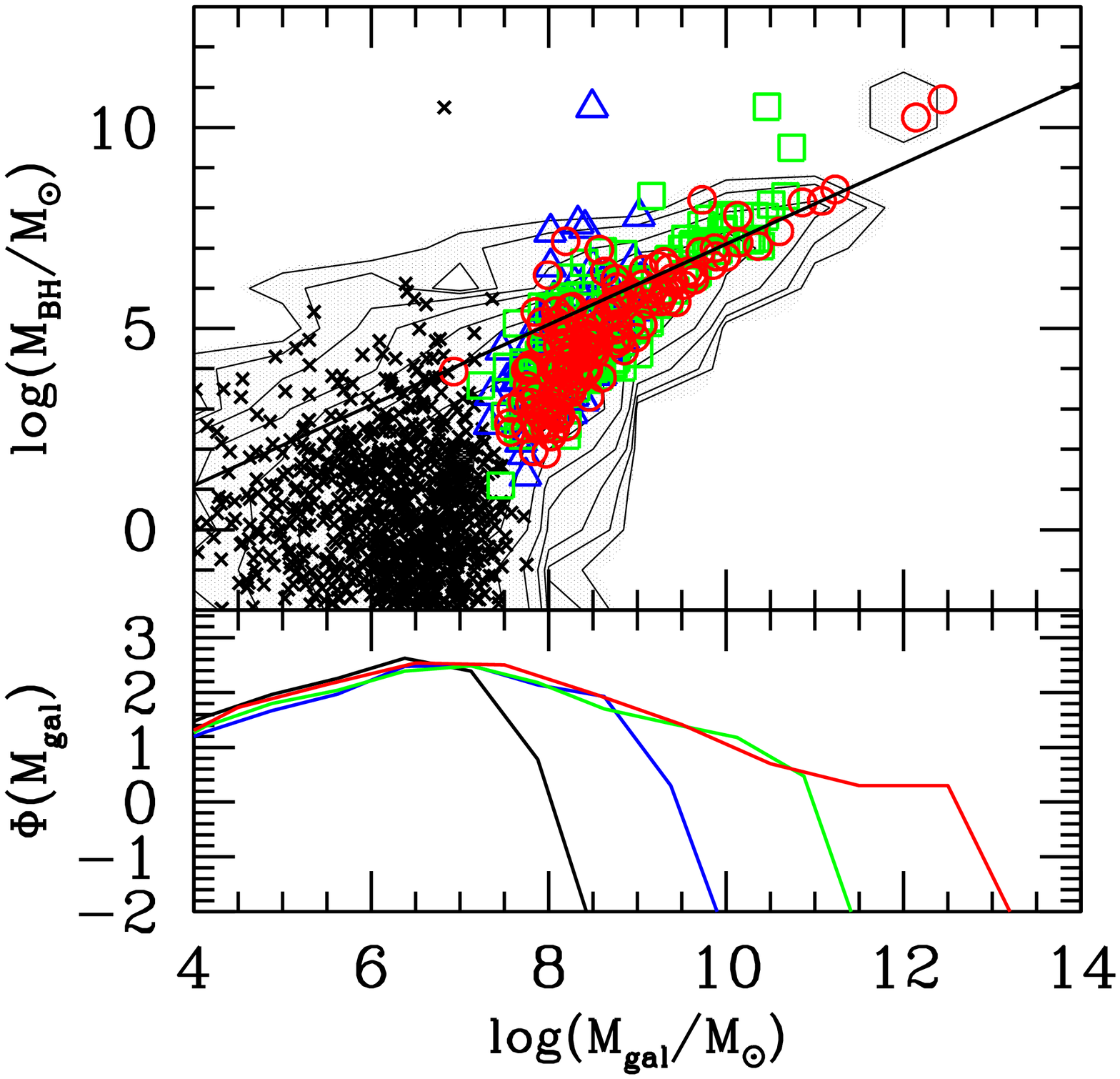}\hfill
                \includegraphics[angle=0,width=9.cm]{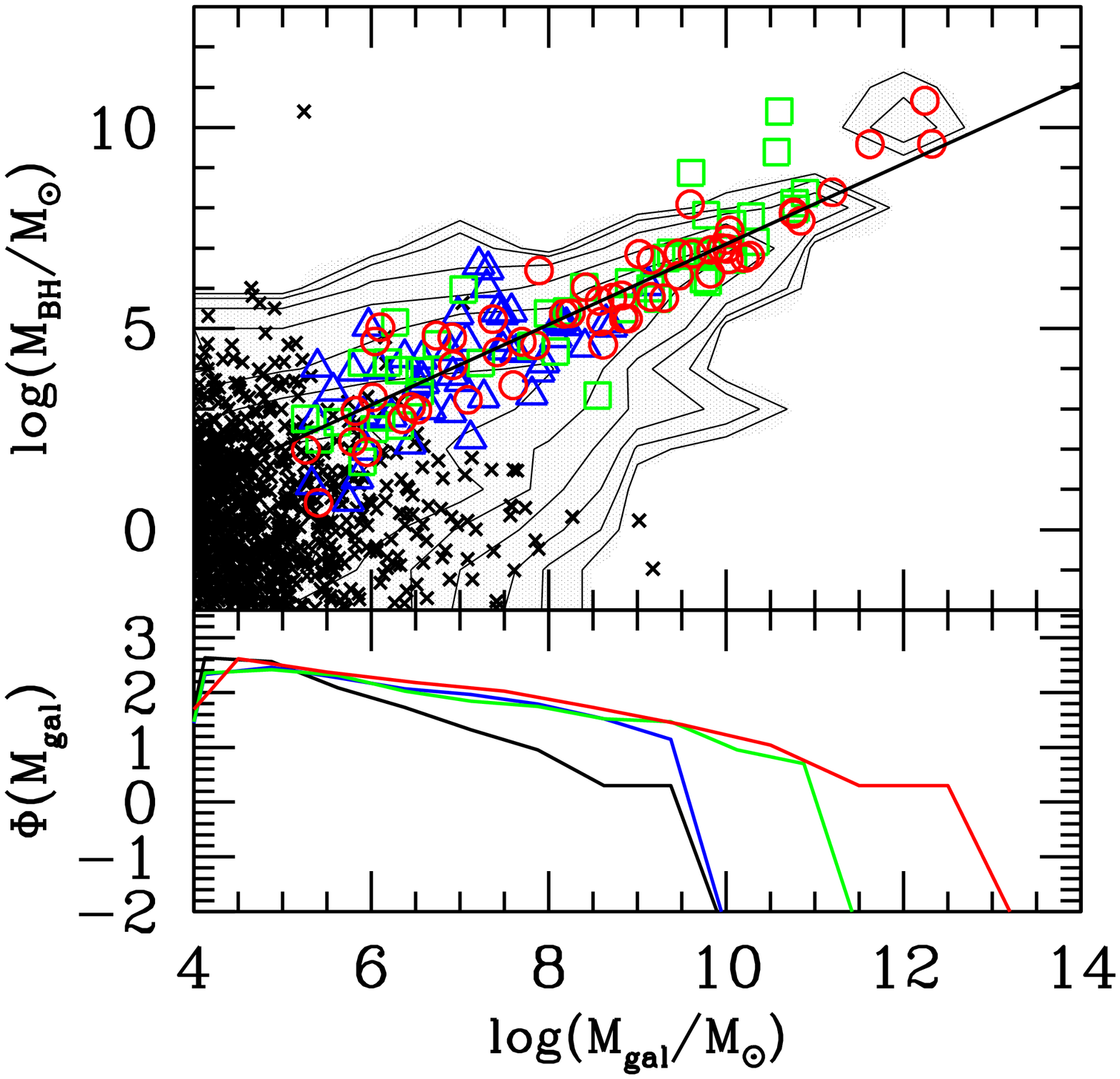}}

    \vspace{0.2in}

    \figcaption{No initial correlation ($\beta=0$) in the \mbh-\mgal\
    relation, in the replenishment scenario.  The black crosses represent the
    initial distribution of points, and the solid line shows the local
    \mbh-\mgal\ relationship from \citet{haering04} -- it is not a fit to the
    data points.  The colored data points represent objects that have
    undergone at least five major merger episodes after 10 ({\it blue
    triangles}), 100 ({\it green squares}), and 1000 ({\it red circles})
    complete ``merging cycles.''  The crosses are the primordial distribution,
    corresponding to the initial mass function.  The shaded region illustrates
    the locus of all points after 1000 cycles; the density of points doubles
    with each contour level.  The cumulative histograms after the
    corresponding merger sequences are shown below the data points.  ({\it a})
    An initial Schechter powerlaw slope of $\alpha=-0.5$.  ({\it b}) An
    initial Schechter powerlaw slope of $\alpha=-1.5$.}
\end{figure*}

\begin{figure*} 
    \label{fig:strongcorr}
    \centerline{\includegraphics[angle=0,width=9.cm]{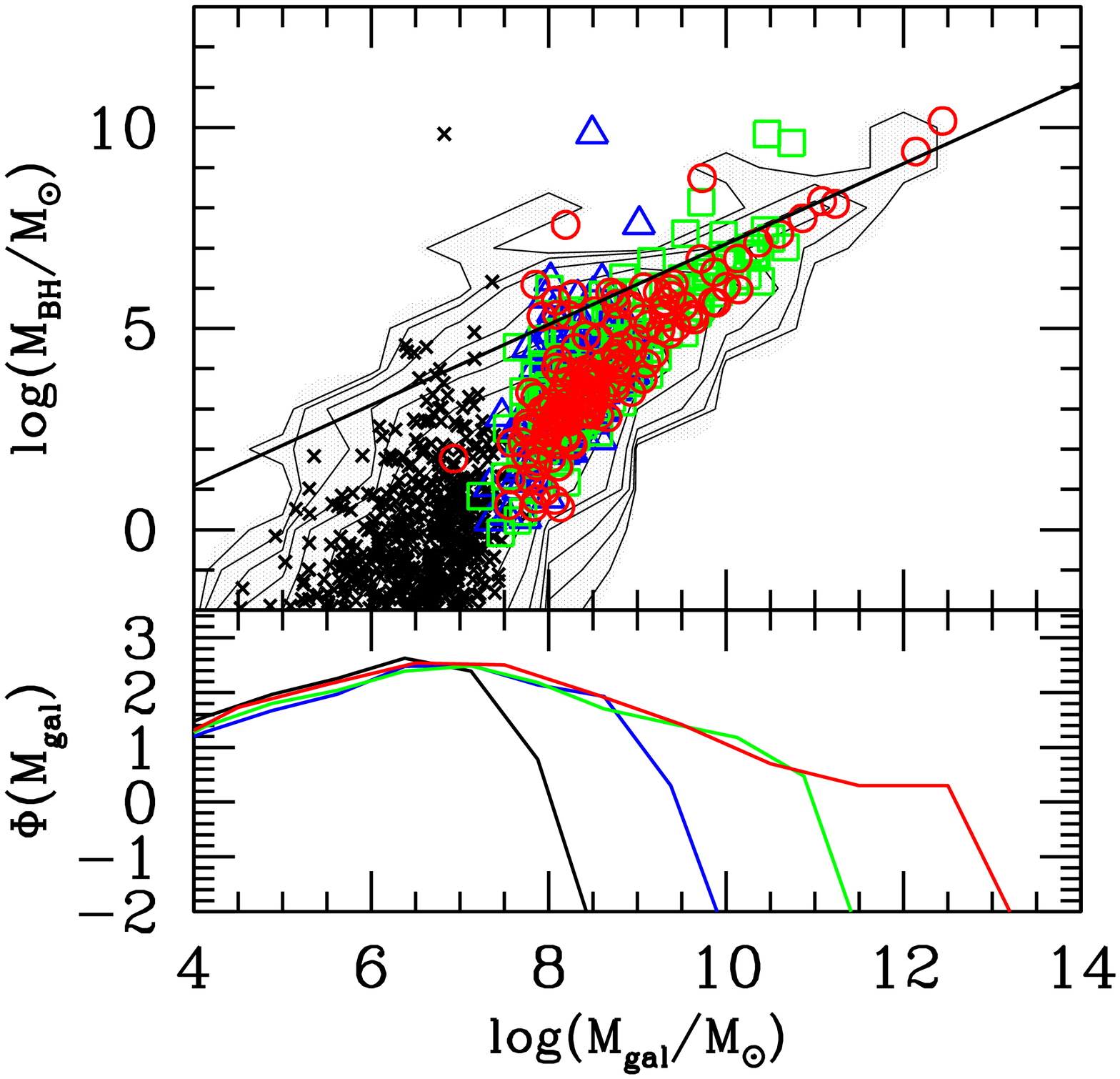}\hfill
                \includegraphics[angle=0,width=9.cm]{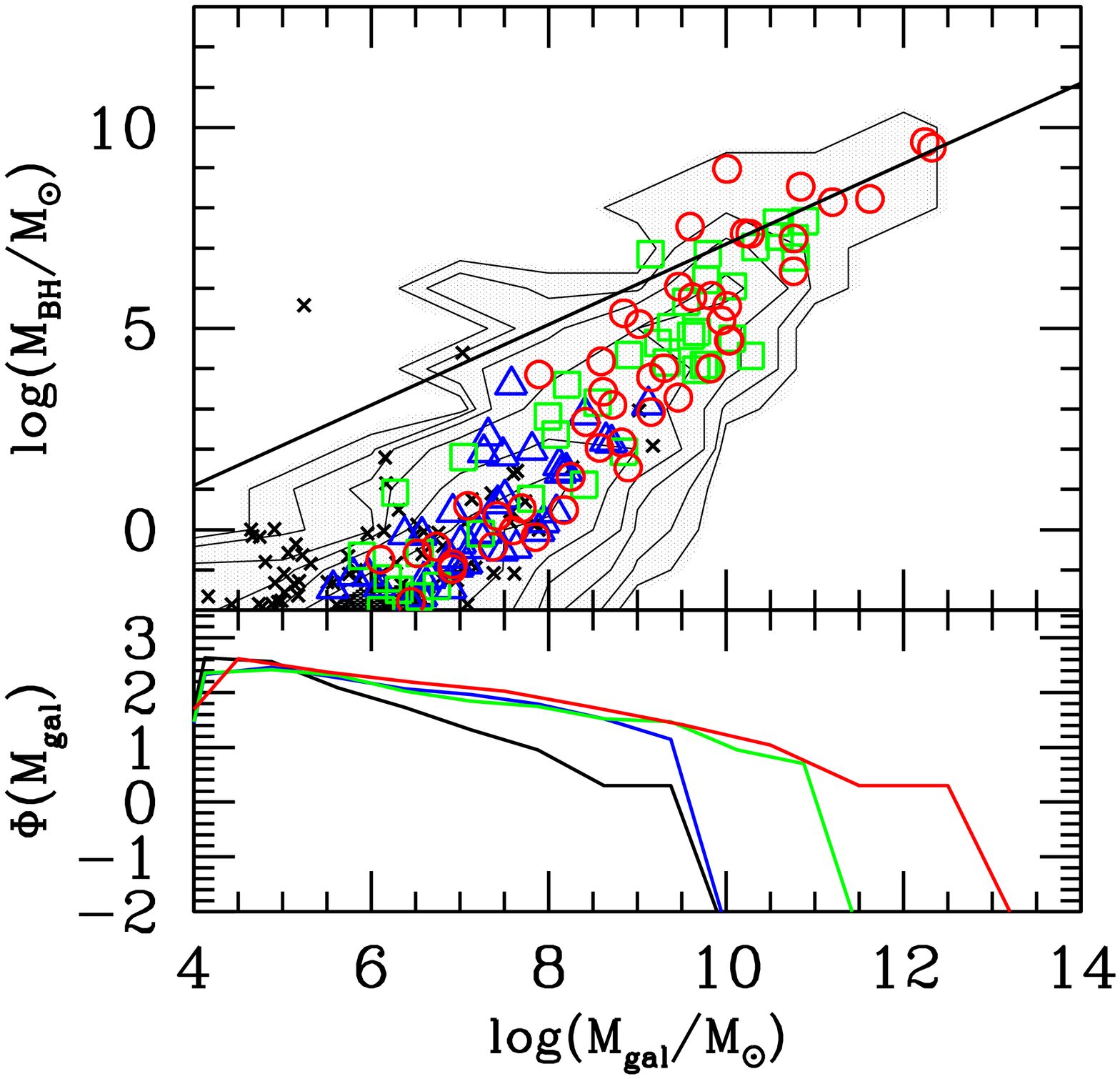}}

    \vspace{0.2in}

    \figcaption{Similar to Figure~3, except for a steep initial correlation
    ($\beta=2$) in the \mbh-\mgal\ relation.  See Figure~3 for details.}
\end{figure*}

\begin{figure} 
    \plotone{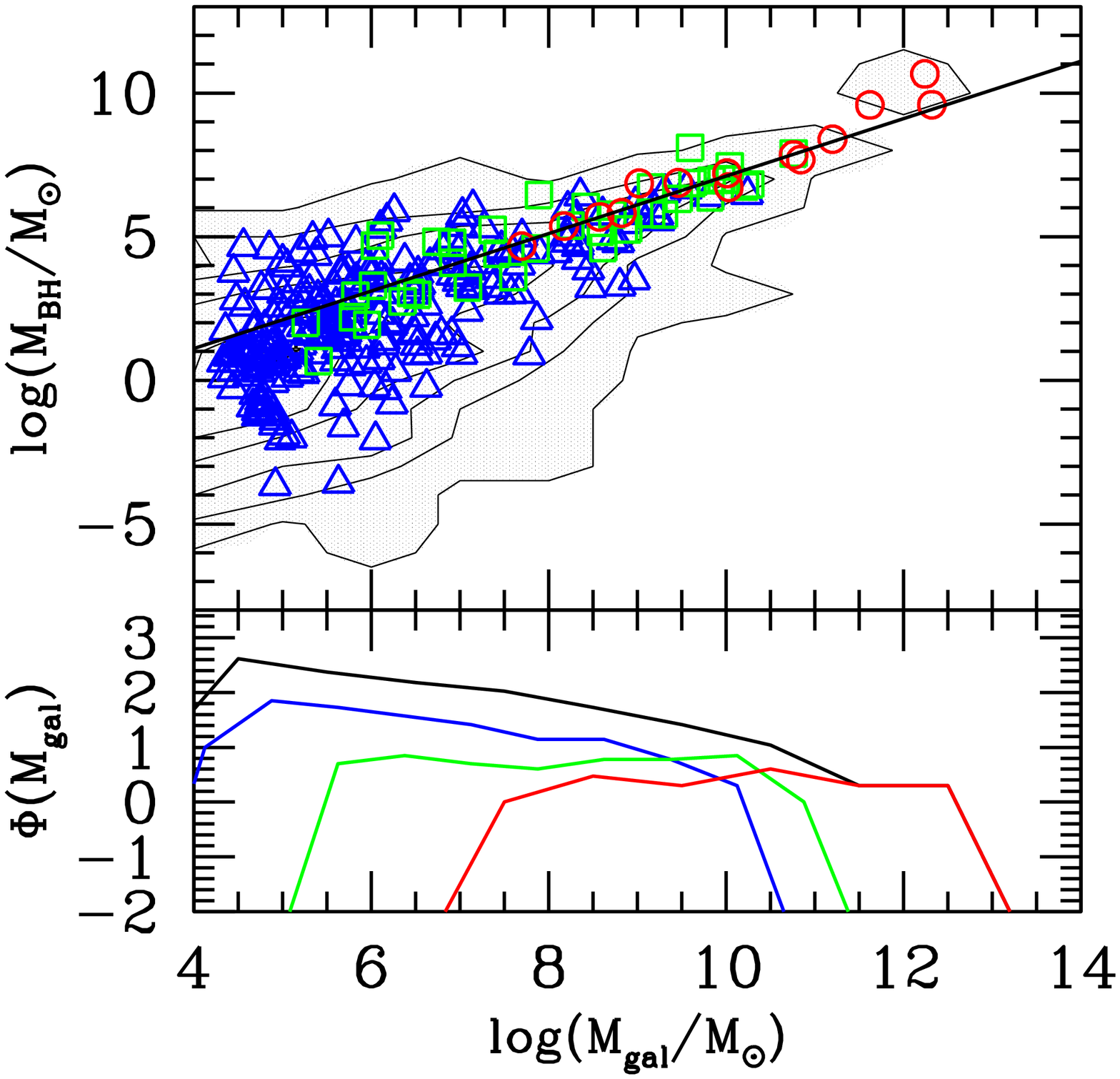} 


    \figcaption{Similar to Figure~3b, showing the effect of central-limit
    tendencies with increasing number of major mergers for galaxies after 1000
    merger cycles.  The colored data points illustrate objects that have
    undergone $1\le N_{\rm{maj}} \le 4$, (blue triangles), $5\le N_{\rm{maj}}
    \le 14$ (green squares), $N_{\rm{maj}} > 14$ (red circles) major mergers.
    The greyscale contours shows the locus of all the points.}
\end{figure}

\begin{figure} 
    \label{fig:nocorr2}

    \plotone{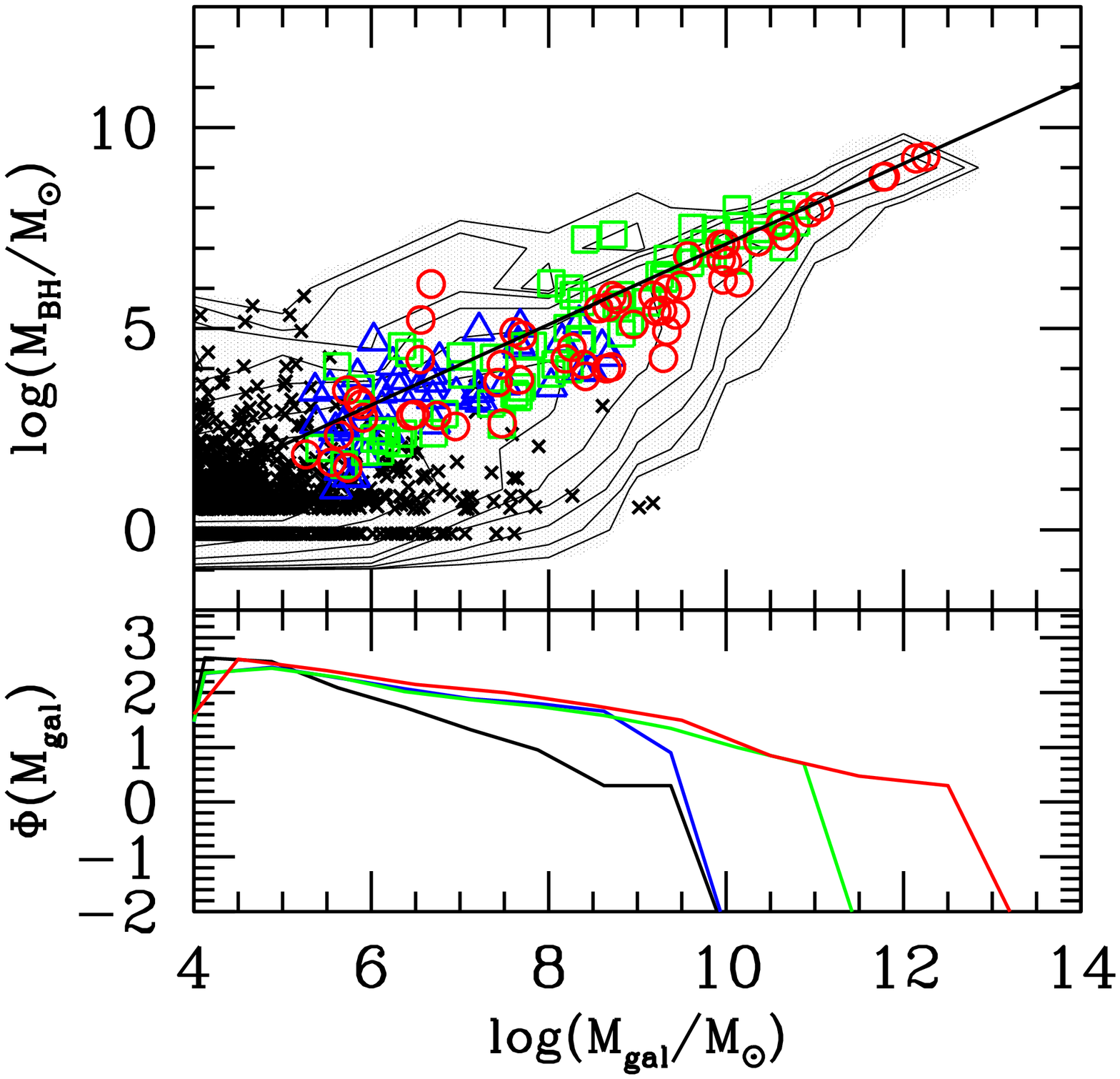}


    \figcaption{Similar to Figure~3, except the BH mass is drawn from a
    Schechter law of $\alpha=-1.5$ instead of a Gaussian distribution.
    See Figure~3 for details.}
\end{figure}

\subsection {Replenishment Scenarios}

The {\it replenishment scenario} is one of the two simple ways considered to
emulate the progress of galaxy evolution.  Here, by definition, the rate of
galaxy mergers equals the rate of galaxy number production.  The way a galaxy
is newly produced is by being selected randomly from an initial mass function,
parameterized by $\alpha$ and \mstar, which does not evolve with time.  In
contrast, galaxies ``grow'' only by merging with another member in the galaxy
pool existing at the time, and hierarchical merging is the only avenue for
mass growth.  Therefore, as galaxies merge, the cumulative mass function does
undergo evolution.  However, the total number of galaxies remains constant
because of the 1:1 ratio of merging:replenishment.

Figures~3 and 4 illustrate the results for initial conditions $\beta=0$ (i.e.
no \mbh-\mgal\ correlation) and $\beta=2$ (steep
\mbh-\mgal\ correlation), respectively, showing only a small subset of the
data points.  In each figure, two different initial mass functions, $\alpha =
-0.5$ (Figs. 3{\it a}, 4{\it a}) and $\alpha=-1.5$ (Figs. 3{\it b}, 4{\it b})
are considered.  In each of the Figures, the initial distribution of the
\mbh-\mgal\ relation (or lack thereof) is shown with crosses.  The open
colored data points show the
\mbh-\mgal\ development of galaxies that have undergone $N_{maj} \ge 5$ major
merger episodes, after 10 (blue triangles), 100 (green squares), and 1000 (red
circles) merger cycles have transpired.  These data points effectively
illustrate the progress of the \mbh-\mgal\ evolution for objects that might be
morphologically identified as early-type galaxies of each cycle.  For clarity,
the contour levels represent the locus of points after 1000 merger cycles, and
the levels are spaced at multiples of 2 in density.  The luminosity functions
of the galaxy pool at the end of the merger cycles are shown in the lower half
of each diagram in corresponding colors and locale in mass.  Lastly, a linear
reference line is overplotted in the Figures with normalization given by
\rn=800 \citep{haering04}, and the simulations are scaled/shifted arbitrarily
to match; it is not a fit to the data points.

As shown in Figures~3 and 4, the convergence towards a tight linear relation
is fairly quick.  After five major merger episodes a linear relation starts to
emerge regardless of the initial conditions of the mass function or the form
of the \mbh-\mgal\ correlation.  One reason for this quick convergence is the
central-limit behavior of major mergers which is shown in Figure 5, in which
the increasing number of mergers is represented by different symbols and
shades.  The one notable case where the convergence toward linearity is slower
than the other scenarios is Figure 4{\it b}, where the effect is only evident
at $10^{10.5}$ \msol\ or greater, even as the scatter has decreased markedly.
In general, if the \mbh-\mgal\ correlation is steep initially, the tail at low
mass remains steep after a large number of major mergers has occurred, even as
the massive end converges toward linearity.

Lastly, the qualitative convergence effects do not depend on the assumption
about the distribution of BH mass at each galaxy mass. Figure~6 shows an
example that is in direct analog to Figure~3b, except that the BH mass is
instead drawn from a Schechter mass function with $\alpha=-1.5$.

\begin{figure*} 
    \centerline{\includegraphics[angle=0,width=9.cm]{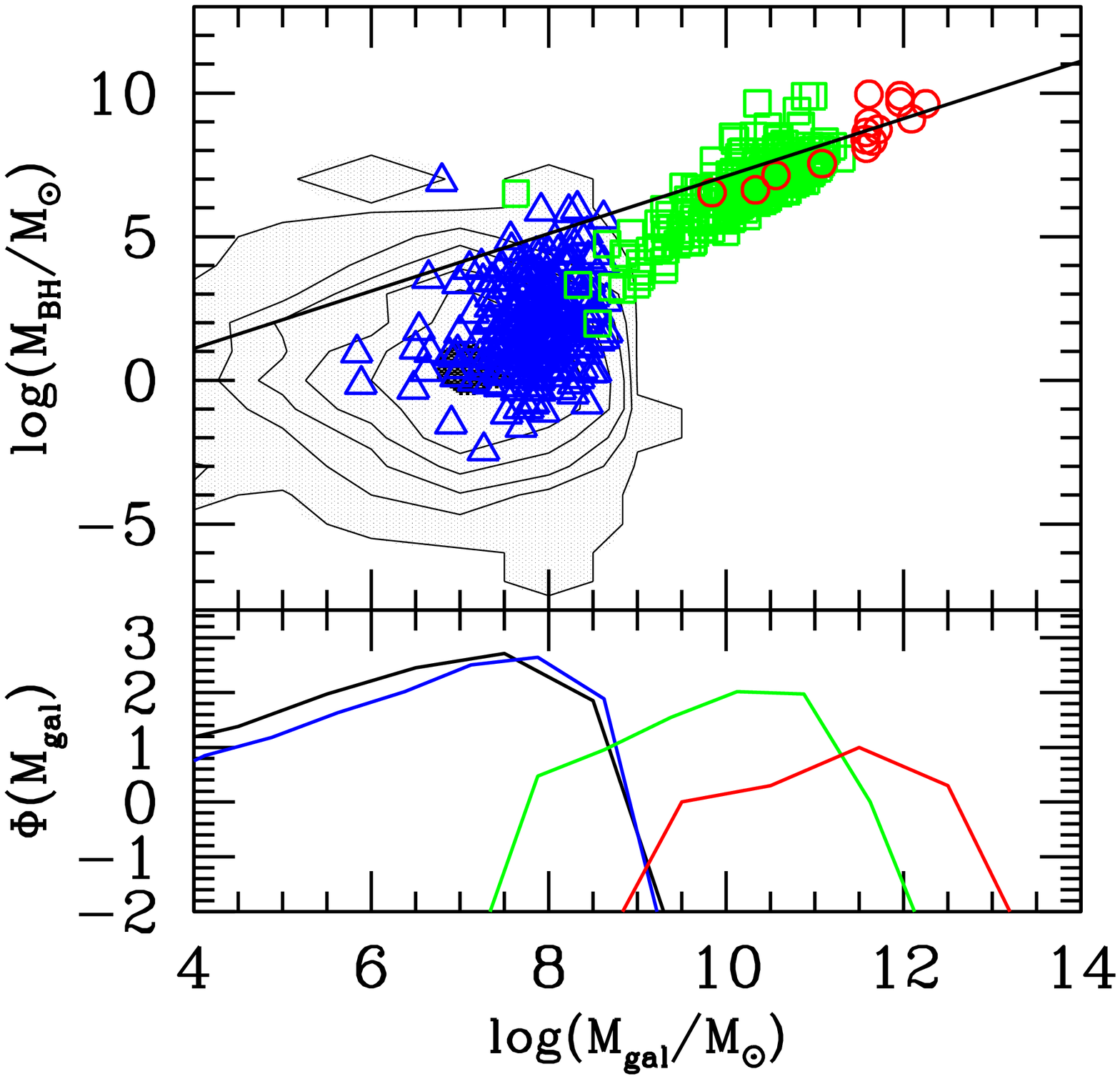}\hfill
                \includegraphics[angle=0,width=9.cm]{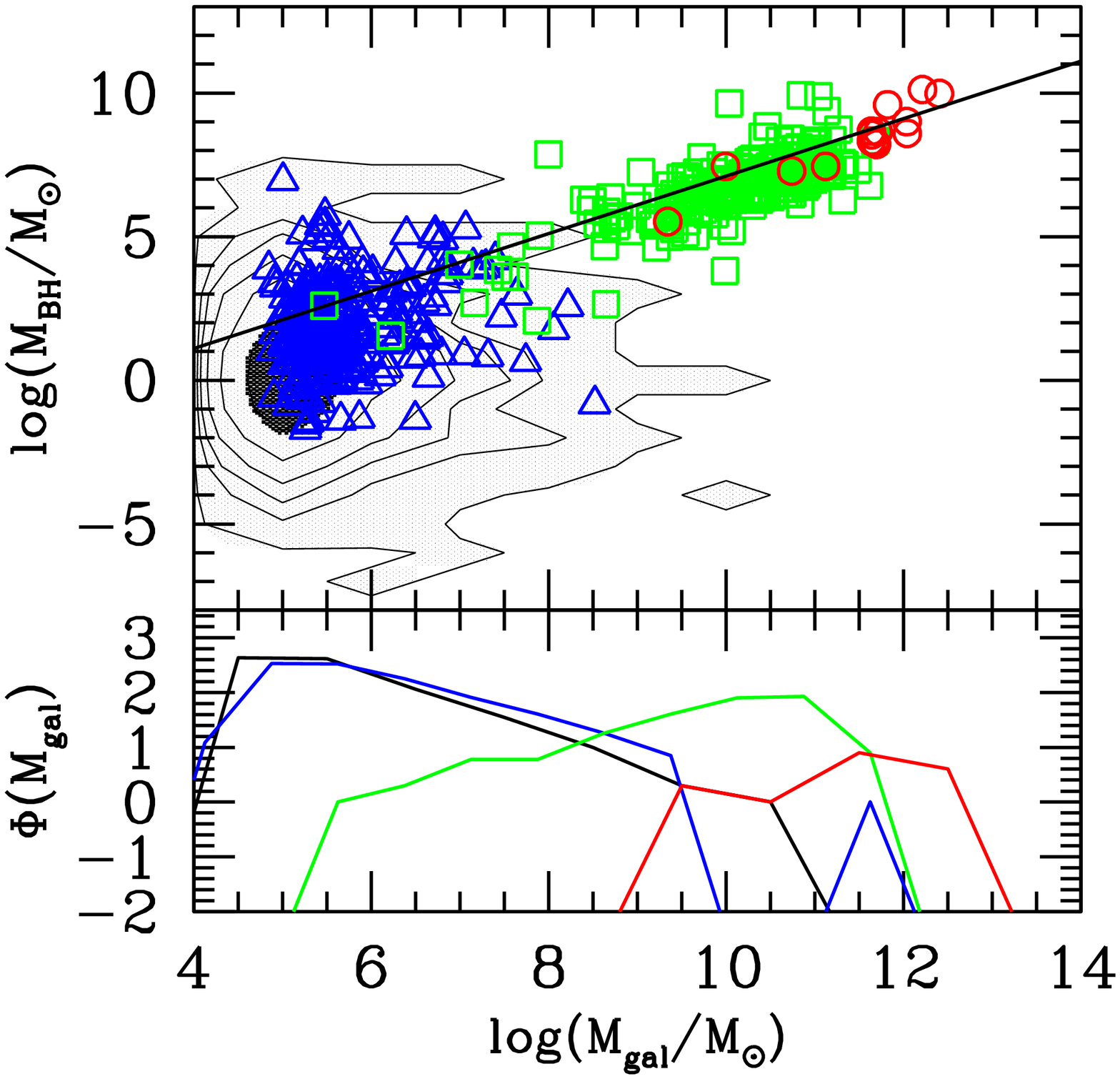}}


    \vspace{0.2in}

    \figcaption{No initial correlation ($\beta=0$) in the \mbh-\mgal\
    relation, depletion scenario.  The contours represent the initial
    distribution of points, and a solid line shows the local \mbh-\mgal\
    relationship from \citet{haering04} -- it is not a fit to the data points.
    The colored data points represent objects that have undergone at least 1
    major merger episodes after 1 complete {\it merging cycles} ({\it blue
    triangles}), 10 ({\it green squares}), and 14 ({\it red circles}).  The
    cumulative histograms after the corresponding merger sequences are shown
    below the data points.  {\it a}) An initial Schechter powerlaw slope of
    $\alpha=-0.5$.  {\it b}) An initial Schechter powerlaw slope of
    $\alpha=-1.5$.}
\end{figure*}

\begin{figure*} 
    \centerline{\includegraphics[angle=0,width=9.cm]{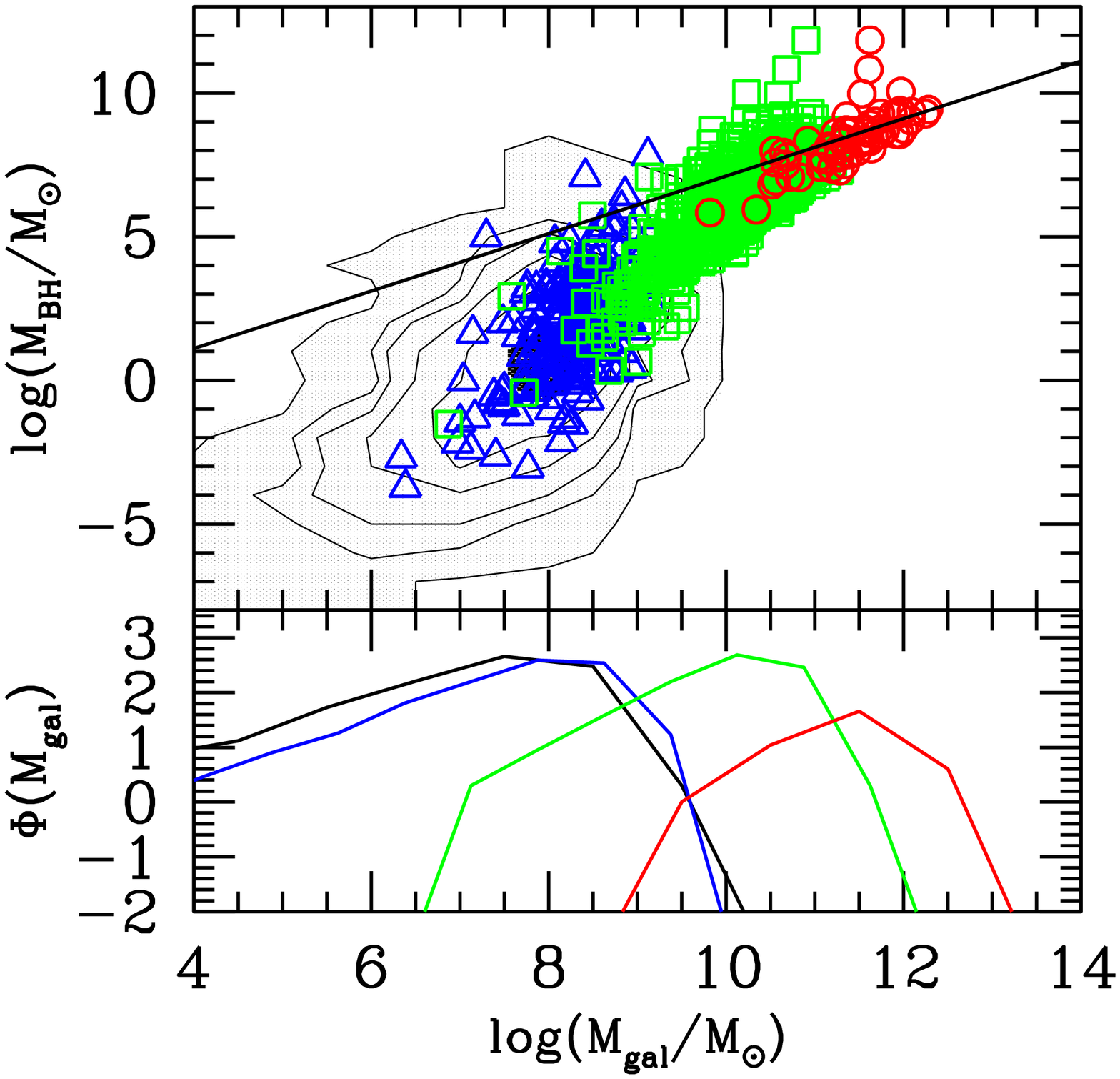}\hfill
                \includegraphics[angle=0,width=9.cm]{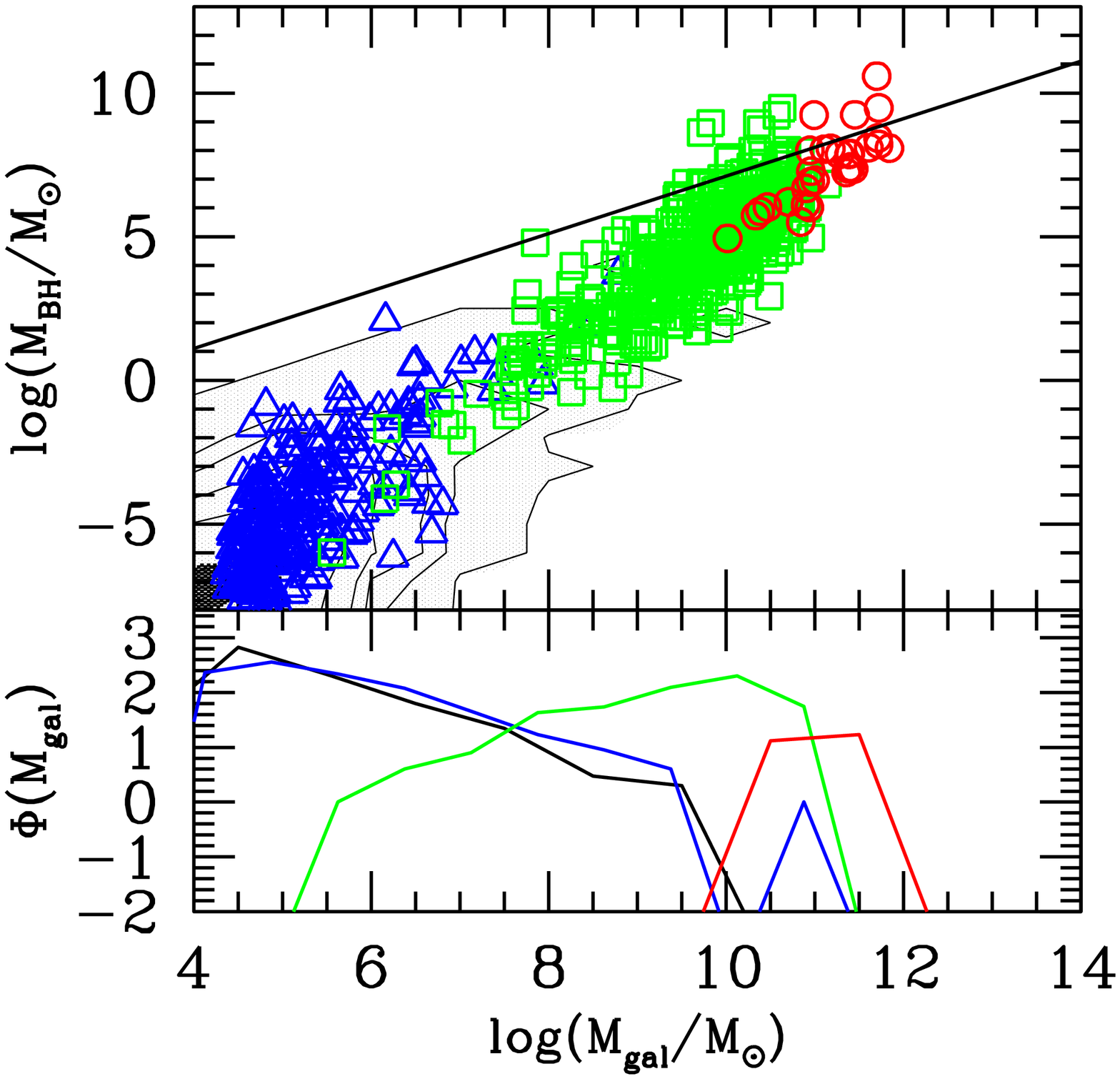}}


    \vspace{0.2in}

    \figcaption{Similar to Figure~7, except for a steep initial correlation
    ($\beta=2$) in the \mbh-\mgal\ relation.  See Figure~7 for details.}
\end{figure*}

    \vspace{0.2in}

\subsection {Depletion Scenarios}

The other extreme of the merger simulations is to consider what effect galaxy
{\it depletion} from a finite reservoir has on the \mbh-\mgal\ relation.
Because the number density of galaxies builds up over time, the depletion
scenario is expected to not be realistic.  Nevertheless, it is useful for
illustrating how the \mbh-\mgal\ convergence is affected by a different
evolution in the mass function as compared with the replenishment scenario.

The depletion scenarios are constructed by creating a large sample of
$5\times10^5$ objects, initially having no correlation between BH and galaxy
masses (Figure~7) or with a $\beta=2$ correlation between the two (Figure~8).
The BH masses are assigned to the galaxies with a log-normal distribution of
dispersion $\sigma=2$ centered around Equation 2.  In each scenario, galaxies
are created to have initial mass functions of $\alpha=-0.5$ (Figures 7a and
8a) or $\alpha=-1.5$ (Figures 7b, 8b).  Then, as galaxies merge, no new ones
are created to replace them.  As a consequence, the mass function evolves by
growing in \mstar, the number density decreases, and a sharp truncation
develops at low masses (see lower half of Figures~7 and 8).  As the number of
merging cycles increases, the scatter decreases quickly and converges toward a
linear relation, as illustrated by the solid line.  Once again, as shown in
Figure~8 (especially 8b), the convergence is much slower for steep $\alpha$
and steep $\beta$ compared with other scenarios.  And while the convergence
trends are noticeable, because of a dearth of minor galaxies with which to
merge at late times (red circles), the slope is virtually ``frozen in,'' and
the subsequent convergence is due mostly to the central-limit theorem.

\section {DISCUSSION AND CONCLUSION}

This study has revisited the issue of how galaxy merging may affect the
\mbh-\mgal\ scaling relation from the standpoint of basic mass addition and
statistics, thereby clearly isolating the merger cause from other detailed
physics that must otherwise affect galaxy evolution.  Through Monte-Carlo
simulations, a tight, linear, \mbh-\mgal\ correlation appears to emerge when
galaxies have undergone five or more major mergers (along the entire tree, not
just the main branch), and many minor ones, for practically all reasonable
initial correlations between \mbh\ and \mgal, or a lack of one.  The main
reasons for these behaviors are seen to be the following:

\begin {enumerate}

\item The galaxy mass function decreases with increasing mass.

\item Major mergers have a strong central-limit tendency, so that regardless
of the initial \mbh-\mgal\ correlation, the {\it scatter} should decrease with
an increasing number of events.  While this tendency also acts on minor
mergers, the drive toward smaller scatter is weaker because minor mergers
occur between galaxies that are vastly discrepant in both galaxy and BH mass
as compared with major mergers, by definition.  The corollary is that the
steeper a correlation between \mbh\ ({\it y}-axis) and \mgal, the stronger the
central-limit tendency for major mergers compared with minor.  However, major
mergers alone are {\it not} enough to cause the \mbh-\mgal\ relation to
converge to {\it linearity} over time because the ratio \mbh/\mgal\ is not
changed much.

\item Minor mergers are primarily responsible for causing the \mbh-\mgal\
relation to converge toward a linear --- that is, \mbh~=~\mratio\mgal ---
relation because the mass function of galaxies follows a Schechter powerlaw.
Without minor mergers, the \mbh-\mbulge\ relation can be ``frozen'' to a slope
that is not necessarily linear.  This linear attractor causes a convergence
toward a tighter \mbh-\mgal\ relation; however, it is less efficient at
reducing the scatter compared with the central-limit seeking tendency of major
mergers, as shown in Figure~5.

\end {enumerate}

It is curious that galaxy merging itself might produce a linear \mbh-\mgal\
relation.  However, a natural question that does arise is, ``When is the
merging statistics presented in this study relevant?''  On the surface, it is
easy to conclude that because the reasoning refers to a two component model it
ought to apply to ``dry'' mergers, but perhaps not to a three component model
involving stars, gas, and BH.  Thus, the implication is also that it ought not
apply to galaxies undergoing gas-rich mergers, that is, early cosmic history.
However, it is not clear that such a skepticism is warranted.  For example, in
the entire discussion thus far, the abscissa, \mgal, might just as well refer
to \mgal=\mstellar+\mgas, instead of just \mstellar.  If BHs do not grow much
by accretion and that the gas does not get removed from the {\it definition}
of \mgal\ during mergers, then the \mbh-\mgal\ correlation can emerge from
statistical merging.  The argument holds true even if \mgas\ transforms
arbitrarily into \mstellar, as long as the sum is conserved.  In the limit
where BHs do grow most of their mass during AGN accretion, as might be implied
by \citet[e.g.][]{soltan82,yu02}, so that $\Delta$\mbh$\propto $\mgas\ and
$\Delta$\mbh $\gg$\mbh, then the correlation between \mbh\ and \mgal\ comes
out {\it by construction} rather than by statistical merging.  However,
statistical reasoning would still be a ``supporting actor'' to reduce the
scatter and to forcibly steer the \mbh-\mgal\ relation in the preferred linear
direction.  Likewise, even if BH growth, or other physics \citep[e.g.
gravitational radiation, three body BH ejection -- ][and references
therein]{merritt04, volonteri05, ciotti07}, were a ``heating'' source, that
is, one that randomizes a tight linear \mbh-\mbulge\ relation, the linear and
central limit attractors would cause a re-convergence if galaxies continue to
merge thereafter by both major and minor mergers.  In summary, while it is
entirely possible that the \mbh-\mbulge\ relation has origins outside of basic
statistics, galaxy merger statistics can still affect the final outcome of a
\mbh-\mgal\ correlation in both the scatter and the slope.  In any event,
statistical reasoning is a fundamentally robust explanation for why random
galaxy merging does not corrupt a pre-existing \mbh-\mbulge\ relation, which
is important to bear in mind in the context of the \mbh-\mbulge\ or
\mbh-\mgal\ relation in a hierarchically forming universe.

While the \mbh-\mbulge\ relation might have other origins, it is nonetheless
interesting and revealing to follow through the consequences of statistical
merging.  For instance, simple statistics naturally explains why black holes
appear to correlate most strongly with galaxy bulges, rather than more
generally with a galaxy as a whole, which might include a stellar disk
\citep{kormendy01}:  bulge masses, assuming they were assembled through major
mergers, have a stronger central-seeking tendency than disk galaxies, whose
growth history might involve more minor merger events.  As such, the
\mbh-\mbulge\ relation is a special case of a more fundamental \mbh-\mgal\
relation.  Reversing the argument, the observational fact that
\mbh\ correlates most strongly with bulge masses, coupled with the
central-limit theorem reasoning, implies that the merger trees of elliptical
galaxies were more dominated by major merger events than were disk galaxies.
Conversely, the fact that the scatter in the \mbh-\mgal\ relation is observed
to be much larger for disk dominated galaxies implies, statistically, that
their progenitors, and progenitors thereof, have undergone more minor mergers.

The possibility that a more fundamental correlation is between \mbh\ and
\mgal\ (rather than \mbulge) also has practical implications for what slope
and scatter would be measured by observations.  First, because the slope
changes with mass even for objects that experienced the same number of major
mergers (e.g.  Figures 3 and 7), the deviation from linearity and the
intrinsic scatter will depend exactly on how the data are cut.  Simply
defining a sample of objects based on a {\it mass} selection cut will bias
one's measurement of the slope and scatter.  Furthermore, defining a sample
based on {\it morphology} criteria may also implicitly preselect samples that
have certain major vs.  minor merger histories.  Observationally, it is
therefore crucial, when comparing intrinsic scatter and slope of the
\mbh-\mbulge\ relation to be specific about sample selection parameter space,
morphology, bulge-to-disk ratios, or other criteria, lest the conclusions be
caused by subtle but trivial selection biases.

Another consequence of this thought experiment is that the ratio
\mratio=\mbh/\mbulge\ approaches an asymptotic value with time from having a
smaller ratio in the past.  On the surface, this appears contrary to the
findings of \citet{peng06b, peng06a,woo06,shields03,shields06b} based on
quasar host galaxy studies that the ratio \mratio\ decreases over time.  If
the quasar host galaxy studies are correct and are not significantly affected
by biases pointed out by \citet{lauer07b}, then some other physics not
considered here is responsible for causing a decline in the normalization of
\mratio\ with time \citep[e.g.  see][]{croton06b, hopkins07a, fontanot06}.
For instance, the abscissa is ambiguous about what mass \mgal\ corresponds.
If gas mass is a significant fraction of a galaxy's mass, then forming stars
out of the gas reservoir would decrease \mratio\ over time, if the abscissa
\mgal\ represents the galaxy's stellar bulge mass.  Secular growth of galaxy
bulges by accreting stars in galaxy disks would also decrease
\mratio, at the expense of increasing the scatter.  Major mergers of pure
stellar bulges, however, would not cause \mratio\ to decrease over time. 

In hindsight, {\it the results} of this study could have been anticipated from
\citet{islam03,islam04,ciotti07}, given that the initial conditions used in
those studies are a special case of this one where the initial BH scatter
$\sigma_{\rm BH} \rightarrow 0$ (Figure 3a or 3b) (M.  Volonteri and L.
Ciotti 2007, private communication).  Just as relevant, \citet{croton06b,
ciotti07} show that once the \mbh-\mbulge\ relation is in place, it is fairly
impervious to being randomized by galaxy merging.  And because \citet{islam03}
also uses realistic cosmological merger trees, they confirm that the arguments
presented here ought to remain relevant.  However, {\it the reasons} behind
the \mbh-\mbulge\ convergence behavior are difficult to extract from previous
studies because of the use of priors, the use of identical BH seeds, the
inclusion of other physics, and the focus on only the BH-bulge coevolution
(i.e. major mergers).  The latter, especially, is worth examining further,
because the prior that one chooses about whether the BH correlates with just
its bulge or with the entire galaxy can lead to differing interpretations.

In particular, one conclusion from \citet{islam03,islam04} is that the
\mbh-\mbulge\ relation converges to a slightly non-linear slope of
$\beta=0.9$; hence they reason that other physics, perhaps BH growth through
accretion of gas, is required in order to increase the slope closer to
linearity.  The reasoning presented in the current study, however, would
stipulate that linearity is an asymptotic outcome of mergers, but deviations
from linearity come from the possibility that the low mass galaxies have not
yet achieved the asymptotic limit, because of a weaker convergence.  At low
masses, the slope deviates from unity in either direction depending on the
initial mass function of the galaxies (e.g. compare Figure 3a with 3b), on the
mass cut of the study, and on the relative incidence of minor versus major
mergers.

An interesting consequence to consider is how the \mbh-\mgal\ relation might
differ between high and low density galaxy clusters.  However, one of the
unrealistic side-effects of using Monte-Carlo simulations to determine merger
rates is that the normalization of the \mbh-\mbulge\ relation is the same in
all density regimes.  This is because the normalization factor, $\Gamma$,
depends only on the ratio of major to minor merger events which, in the
Monte-Carlo universe, is not affected by a simple rescaling of the mass
function.  However, in the real universe, the relative rates of major and
minor mergers can change with density and, as such, may result in different
normalization and scatter in the \mbh-\mgal\ relation.

Lastly, because of the ambiguity in what \mgal\ corresponds, depending on
whether it refers to the total stellar mass, gas mass, dark matter halo mass,
or a combination thereof, the degree of scatter and linearity would clearly
differ, as a result of different initial mass functions and merger histories.
Because the scenarios considered above depend on a linear addition of masses,
the arguments therefore may not apply to gas masses that are not
gravitationally bound to a galaxy.  These and other issues will be addressed
in a future study, which will incorporate the use of realistic merger trees.

\section*{Acknowledgment}

Over the course of this study, I have greatly enjoyed lively discussions with
many friends and colleagues, including Jenny Greene, Hans-Walter Rix, Luis Ho,
Eric Bell, Michael Santos, Robert Kennicutt, Rachel Somerville, Aaron Barth,
Christy Tremonti, Darren Croton, Tommaso Treu, and David Koo.  I also thank
Jim Rose, Wayne Christiansen, Marta Volonteri, Luca Ciotti, and Alister Graham
for comments and discussions on past and future studies.  This study greatly
benefited from discussions with Phil Hopkins on issues related to AGN feedback
over the past several months.  I also gratefully acknowledge insightful
comments and suggestions from the referee, and Science Editor Chung-Pei Ma,
and the support of STScI through the Institute (Giacconi) Fellowship Program.

\begin{appendix}

This appendix shows that the \mbh-\mbulge\ relation follows a
central-limit-like behavior when galaxies undergo major mergers.
Specifically, this means that if the initial parent distribution of progenitor
\mbh\ that undergoes merging is, for simplicity, normally distributed about a
mean BH mass $\mu$, thus having a logarithmic dispersion
$\sigma(\mbox{log}(\mu_{\rm{init}}))$, then a new distribution of {\mbh}s
after merging will have a log-normal dispersion that scales as:
$\sigma\left(\mbox{log}\left(\mbox{\mbhmerge}\right)\right)
\sim \sigma(\mbox{log}(\mu_{\rm{init}})) / \sqrt{2}$ .

\medskip

First, the mean, $\mu_{\rm{merge}}$, of the resulting BH distribution after
two BHs merge from the initial parent distribution is:

\begin {equation}
    \mbox{\mbhmerge} = \left<\mbox{\mbh$_{,1}$} + \mbox{\mbh$_{,2}$}\right>,
\end {equation}

\noindent \noindent where \mbh$_{,1}$ and \mbh$_{,2}$ are drawn from the same
parent distribution for major galaxy mergers.  Then,

\begin {equation}
    \mbox{log}\left(\mbox{\mbhmerge}\right) = \mbox{log}\left<\mbox{\mbh$_{,1}$} + \mbox{\mbh$_{,2}$}\right>,
\end {equation}

\noindent From propagation of errors the log-normal error is:  $\sigma
(\mbox{log}(x)) = \sigma(x)/x$, then,

\begin {equation}
    \sigma\left(\mbox{log}\left(\mbox{\mbhmerge}\right)\right) =
	\frac{\sigma\left<\mbox{\mbh$_{,1}$} + 
	\mbox{\mbh$_{,2}$}\right>}{\left<\mbox{\mbh$_{,1}$} + \mbox{\mbh$_{,2}$}\right>},
\end {equation}

\noindent By definition, a distribution obtained by averaging the mass of
merging BH pairs is a normal distribution with a mean of the initial parent
distribution, $\mu_{\rm{init}}$:

\begin {equation}
    \mu_{\rm{init}} = \frac{\left<\mbox{\mbh$_{,1}$} + \mbox{\mbh$_{,2}$}\right>}{2} 
        = \frac {\mbox{\mbhmerge}}{2}.
\end{equation}

\noindent Because \mass$_{,1}$ and \mass$_{,2}$ are drawn from a normalized
distribution around a parent mean $\mu$, the new distribution of
$\sigma(\mu_{\rm{merge}}) \equiv \sigma\left<\mbox{\mass$_{,1}$ + \mass$_{,2}$}\right>$ is:

\begin {equation}
    \sigma \left(\mbox{\mbhmerge}\right) \sim \frac{\sigma(\mu_{\rm{init}})}{\sqrt{2}} \times 2.
\end {equation}

\noindent Substituting Eqs. A4 and A5 into A3 yields:

\begin {equation}
    \sigma\left(\mbox{log(\mbhmerge)}\right) \sim \frac{\sigma(\mu_{\rm{init}})}{\mu_{\rm{init}}\sqrt{2}}.
\end {equation}

\noindent  Using the fact that, $\sigma (\mbox{log}(\mu)) = \sigma(\mu)/\mu$,
Equation A6 becomes:

\begin {equation}
    \sigma\left(\mbox{log(\mbhmerge)}\right) \sim 
		\frac{\sigma\left(\mbox{log}(\mu_{\rm{init}})\right)}{\sqrt{2}}. 
\end {equation}

\end {appendix}

\bibliography{references}

\end {document}